\documentclass[pre,a4,twocolumn,notitlepage,superscriptaddress,color,showpacs]{revtex4-2}
\usepackage{amsmath,amsfonts,amssymb,float}
\usepackage[pdftex]{graphicx}
\usepackage{xcolor,comment}

\begin{document}
\title{Giant strongly biconnected components of directed networks: a generating function approach}

\author{Minsoo Yang}
\affiliation{Department of Physics, Chungbuk National University, Cheongju, Chungbuk 28644, Korea}
\author{Reinhard Laubenbacher} 
\email{reinhard.laubenbacher@medicine.ufl.edu}
\affiliation{Department of Medicine, University of Florida, Gainesville, Florida 32610, USA}
\author{Byungjoon Min}
\email{bmin@cbnu.ac.kr}
\affiliation{Department of Physics, Chungbuk National University, Cheongju, Chungbuk 28644, Korea}
\affiliation{Department of Medicine, University of Florida, Gainesville, Florida 32610, USA}
\affiliation{Advanced-Basic-Convergence Research Institute, Chungbuk National University, Cheongju, Chungbuk 28644, Korea}
\date{\today}\date{\today}

\begin{abstract}
Strongly connected components (SCCs) characterize modular structure in directed networks but are fragile to single node failures. 
We study strongly biconnected components (SBCs), which are the set of nodes in which every node pair remains mutually reachable after the removal of any single node, as a more robust notion of connectivity. 
Using a generating function formalism, we derive the size of the giant SBC and analyze its percolation behavior under random node and link removal. 
We show that the giant SBC emerges at the same threshold as the giant SCC but grows more slowly, consistent with previous findings for undirected networks.
We also applied our theoretical framework to real-world biological networks including gene regulatory networks and neural connectome.
Our framework provides insight into the interplay between connectivity, redundancy, and robustness in complex directed systems.
\end{abstract}
\maketitle

\section{Introduction}

Structural robustness of networked systems plays an important role 
in understanding the integrity and function of complex systems, 
including biological, technological, and social systems \cite{albert2000error,artime2024robustness,cohen2000resilience}. 
To this end, the robustness of networks has long been a central problem 
in statistical physics, network science, and complexity science
\cite{cohen2000resilience,callaway2000network,cohen2001breakdown,schneider2011mitigation,min2014network}. 
The largest connected component is central to the robustness and 
functioning of these systems at a global scale 
\cite{albert2000error,molloy1995critical,buldyrev2010catastrophic,jang2026feedback}. 
In this regard, the size of the giant component has become the basic 
measure of the robustness of complex networks, which is determined by the connectivity patterns of the networks \cite{albert2000error,cohen2000resilience,cohen2001breakdown,artime2024robustness}. 
Therefore, the formation of the giant component and its response to the removal of elements has received much attention, with a direct connection to percolation theory \cite{stauffer2018}.

In undirected networks, connected components are the basic units of network structure.
However, many real-world systems are inherently directed, where the direction of interactions 
carries essential functional meaning \cite{garlaschelli2004patterns, dorogovtsev2001giant,leicht2008community,son2009dynamics,kim2010finding,marmulla2024centrality,bianconi2008local}. 
In directed networks, connected paths need to be defined more specifically, 
and hence the building blocks of connected patterns are commonly characterized by 
strongly connected components (SCCs) \cite{nuutila1994finding,dorogovtsev2001giant,
boguna2005generalized,timar2017mapping,kadelka2023modularity,Niall2023strong,liu2026optimal}
An SCC is a maximal set of nodes in which every node is reachable 
from every other node via directed paths in both directions. 
The existence of such mutual reachability underlies the emergence of 
cyclic loop and coordinated function within network modules \cite{nuutila1994finding,dorogovtsev2001giant,kadelka2023modularity,wu2010feasibility}.
Despite their utility as functional modules, SCCs are inherently fragile: 
the removal of even a single node can disconnect an SCC and destroy reachability between pairs within the SCC \cite{kadelka2023modularity,wu2010feasibility}. 
Therefore, a more robust notion of connectivity is needed. 
Specifically, one may ask which node pairs remain mutually reachable under any single node failure.

In this paper, we analyze the biconnectivity in directed graphs,
focusing on the size of the giant strongly biconnected component (SBC) as a framework for characterizing robust connectivity, in order to overcome the vulnerability of SCC \cite{newman2008bicomponents,georgiadis2016,georgiadis2018,jaberi2016computing,mohseni2021percolation}.
A pair of nodes $i$ and $j$ is said to be strongly biconnected 
if they remain mutually reachable following the removal of any single intermediate node. 
This condition is equivalent to the existence of at least two node-disjoint paths 
from $i$ to $j$ and from $j$ to $i$ \cite{newman2008bicomponents,georgiadis2016}.
An SBC is then defined as a maximal set of nodes in which every pair satisfies this condition. 
By definition, all nodes in an SBC remain within the same SCC under any single node removal, 
rendering the SBC a natural measure of the stable block of a directed network. 
This generalizes the notion of biconnectivity \cite{newman2008bicomponents,azimi2013core,azimi2014giant} 
from undirected to directed graphs. 
We adopt the term ``strongly biconnected component'' or strongly bicomponent in short to emphasize the parallel with the undirected 
bicomponent \cite{newman2008bicomponents}.
Note that the paths between a pair in an SBC may contain nodes outside the SBC; 
thus SBCs need not be self-contained subgraphs, which is analogous to the behavior of 
$k$-components with $k \geq 3$ in undirected graphs \cite{newman2008bicomponents}. 
In the algorithmic graph theory literature, SBCs correspond to the 2-vertex-connected blocks 
of directed graphs \cite{georgiadis2016,georgiadis2018}. 
Other related studies have studied the size of the $k$-core \cite{zhao2017generalized}, 
giant component on multiplex directed networks \cite{azimi2014giant}, and 
core organization \cite{azimi2013core} of directed networks.

Another motivation for this work comes from the modular organization of biological systems, 
which has been a central topic in systems biology since the introduction of general systems theory. 
Biological systems are widely believed to consist of highly cohesive modules that are loosely connected to one another \cite{hartwell1999molecular,ravasz2002hierarchical}. 
Despite extensive research on biological modularity, developing a rigorous theoretical framework that can identify modules and relate their structural organization to the dynamics and control of biological systems remains an important challenge.
Recently, a theoretical framework for modularity in biological Boolean networks has been proposed, in which modules are defined as SCCs of the underlying dependency graph \cite{kadelka2023modularity}. 
This definition is motivated by the importance of feedback loops in biological dynamics and provides a structural decomposition that can be used to analyze and control biological systems. 
However, the decomposition of SCCs is sensitive to changes in network connectivity because the addition or removal of even a single directed link may significantly alter the modular structure. 
Moreover, biological network models may contain uncertain interactions or feedback loops.

While the theoretical framework for biconnectivity in undirected graphs has been developed \cite{newman2008bicomponents,kim2013phase,kim2015biconnectivity}, 
the analogous theory for directed networks has received little attention. 
Previous studies on undirected networks have shown that the giant bicomponent emerges at the same percolation threshold as the giant connected component but grows more slowly above the threshold \cite{newman2008bicomponents,kim2013phase}. 
In directed networks, however, biconnectivity requires independent directed paths in both directions, and its behavior remains largely unexplored.
We here address this gap by proposing an analytical framework to determine 
the size of the giant SBC in directed networks using the generating function formalism based on percolation theory \cite{newman2001random,newman2008bicomponents,dorogovtsev2001giant}. 
In this work, we confirm our theory using numerical simulations in directed random and real-world networks. 
We further extend the framework to characterize the robustness of the giant SBC under 
random failures and apply it to biological networks to assess their structural robustness.

The remainder of this paper is organized as follows. 
In Sec.~2, we define strongly biconnected components and present their importance.
In Sec.~3, we develop the analytical framework based on generating functions for  computing the size of the giant SBC and the percolation threshold 
at which the giant SBC emerges. 
In Sec.~4, we apply this framework to random directed graphs 
and further apply the approach to real-world biological networks. 
Finally, Sec.~5 discusses the implications of our findings and outlines directions for future work.

\section{Strongly biconnected components}

\begin{figure}
\includegraphics[width=0.8\linewidth]{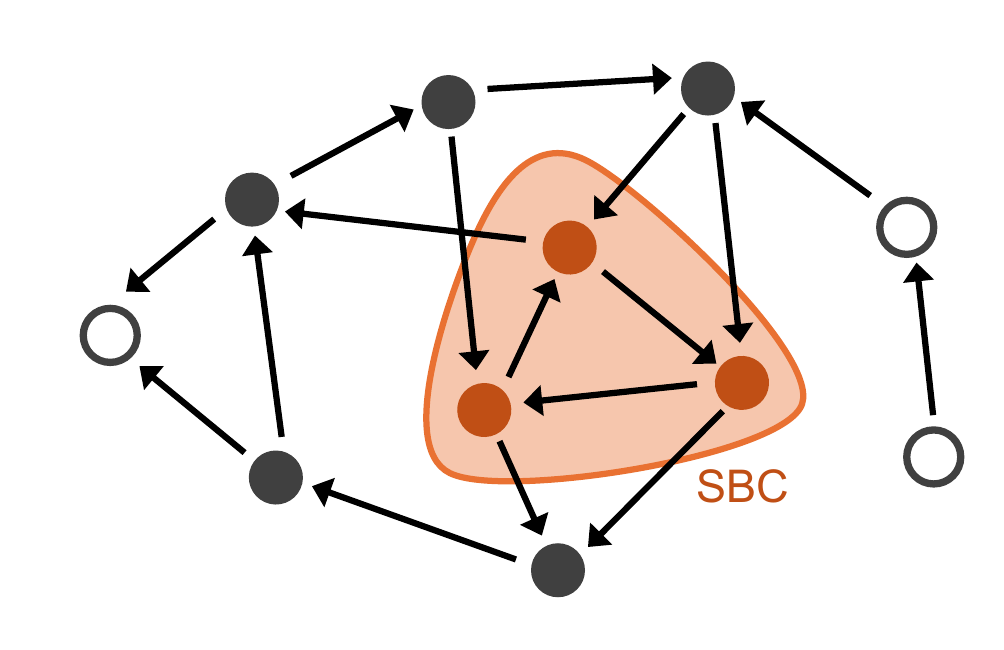}
\caption{
Schematic example of strongly bicomponents that are defined as 
the set of nodes that are connected by at least two disjoint paths.
Filled nodes represent the strongly connected component (SCC), 
while the orange-shaded region highlights the SBC. 
Open symbols denote non-SCC nodes.
}
\label{fig:fig1}
\end{figure}

Directed networks differ from undirected ones: a directed link from node $i$
to $j$ does not imply the existence of a return link from $j$ to $i$. 
As a result, paths in directed networks are not symmetric, and therefore the concept of a connected component 
must be defined in terms of directed reachability \cite{dorogovtsev2001giant}.
In this regard, a unit of connectivity in directed networks is the strongly connected component, 
a maximal set of nodes in which every node can reach every other node via directed paths in both directions.
However, SCCs are inherently fragile: the removal of a single node in a SCC can destroy the reachability.
In practice, since noise and errors are unavoidable in many real-world networks, such vulnerability raises concerns 
for identifying the robustness and functionality of complex networks.

This leads to the need for a more robust notion of connectivity in directed networks.
We define a pair of nodes $(i,j)$ to be strongly biconnected if it remains within the same SCC after the removal of any single intermediate node. 
To satisfy this condition, there must exist at least two node-disjoint directed paths from $i$ to $j$ and from $j$ to $i$. 
A maximal set of nodes satisfying this condition for all pairs is defined as 
a strongly biconnected component (Fig.~1). 
This concept requires not merely the existence of directed paths, but the redundancy against node failures. 
In the algorithmic graph theory literature, SBCs correspond to the 2-vertex-connected blocks of directed graphs \cite{georgiadis2016,georgiadis2018}.
While the giant SCC is a measure of global connectivity in directed networks,
the giant SBC identifies the robust core embedded within the SCC. 
This distinction is particularly relevant in real-world directed networks where node failures are unavoidable.

Understanding when and how emerging the giant SBC provides 
insight into the structural resilience of directed networks. 
In biological networks such as gene regulatory circuits or neural systems, 
the persistence of coordinated function under perturbation is essential, 
and an SBC guarantees that any two nodes within it maintain mutual reachability after the loss of any single intermediate node. 
The emergence of the giant SBC is rooted in the concept of structural redundancy, that is, 
the existence of multiple independent paths between nodes. 
While such redundancy provides an advantage by ensuring system resilience against node failures, 
it can lead to a trade-off since maintaining redundant links generates physical, metabolic, or economic costs. 
Therefore, analyzing the giant SBC not only quantifies a network's robustness but also 
provides insight to how real-world systems optimize the inherent trade-off between 
the cost of wiring and the necessity of fault tolerance.

\section{An Analytical Approach Based on Generating Function Methods}

In this section, we present a theoretical framework using the generating function formalism to quantify the size of the giant SBC. 
We first recast the existing theory for the size of the giant SCC.
We then generalize it to derive equations governing the size of 
the giant SBC and estimate the percolation threshold at which the giant SBC first emerges. 
Finally, we extend this framework to analyze how the size of the giant SBC changes under node and link failures.
It is worth noting that our generating function approach is based on the directed configuration model, in which links are randomly assigned according to a prescribed joint in- and out-degree distribution $P(k_i,k_o)$.
This framework allows correlations between the in-degree and out-degree of individual nodes but does not account for degree correlations between connected nodes or other structural correlations.

\subsection{Giant strongly connected component}

To compare the behaviors of the giant SCC and SBC, we derive the size of the giant SCC in a directed network. 
Consider the joint degree distribution $P(k_i, k_o)$ of the directed network, where $k_i$ and $k_o$ represent the in-degree and out-degree, respectively. 
We define the generating function for this distribution as \cite{dorogovtsev2001giant,newman2001random,kryven2017general}
\begin{align}
G(x,y) = \sum_{k_i,k_o} P(k_i,k_o) x^{k_i} y^{k_o}.
\end{align}
Next, we introduce two generating functions associated with the in- and out-components by following directed edges against and along their directions, respectively \cite{dorogovtsev2001giant}:
\begin{align}
G_i(x,y) &= \sum_{k_i,k_o} \frac{k_o P(k_i,k_o)}{z} x^{k_i} y^{k_o - 1},\\ 
G_o(x,y) &= \sum_{k_i,k_o} \frac{k_i P(k_i,k_o)}{z} x^{k_i - 1} y^{k_o},
\end{align}
where $z$ denotes the mean degree, with the average in-degree and out-degree both equal to $z$.
Here, $G_i(x,y)$ describes the degree distribution of a node reached by following an incoming edge against its direction. 
Conversely, $G_o(x,y)$ describes the degree distribution of a node reached by following an outgoing edge along its direction.
We further define giant in and out-component. The giant in-component is the set of nodes from which the giant SCC can be reached through directed paths, whereas the giant out-component is the set of nodes that can be reached from the giant SCC through directed paths.

On locally tree-like networks, we can compute the probabilities $u$ and $v$ which represent the probabilities that a node arrived at by following 
a randomly chosen in-link or out-link does not lead to the giant in-component or giant out-component \cite{dorogovtsev2001giant}:  
\begin{align}
u &= G_i(u,1), \\
v &= G_o(1,v).
\label{eq:uv}
\end{align}
The probability that a randomly selected node belongs to the giant SCC is then  
the probability that a node is connected to both the giant in- and out-components through at least one in-link and one out-link. 
Thus, the size of the giant SCC on locally tree-like structures is \cite{dorogovtsev2001giant}
\begin{align}
S &= \sum_{k_i,k_o} P(k_i,k_o) \left[ (1 - u^{k_i})(1-v^{k_o}) \right] \nonumber \\
&= 1 - G(u,1) - G(1,v) + G(u,v).
\end{align}
We further define the marginal in-degree and out-degree distributions
as $P(k_i) = \sum_{k_o} P(k_i,k_o)$ and $P(k_o) = \sum_{k_i} P(k_i,k_o)$, respectively.
When the in- and out-degree distributions are uncorrelated, i.e., 
$P(k_i,k_o)=P(k_i)P(k_o)$, the expression simplifies to
\begin{align}
S &= [1 - G(u,1)][1 - G(1,v)].
\end{align}
If the in- and out-degree distributions are further assumed to be identical, this reduces to
\begin{align}
S &= [1 - G(u,1)]^2.
\end{align}

\subsection{Giant strongly bicomponent}

Following the approach to biconnectivity in undirected networks [26], we consider locally tree-like directed random networks in the thermodynamic limit. 
In this limit, we assume that distinct branches leading to the giant in- and out-components provide independent paths through the giant strongly connected structure. 
Accordingly, we neglect configurations in which multiple connections to the giant in- and out-components do not provide the required independent directed paths between nodes. 
In other words, such configurations are assumed to make a negligible contribution to the size of the giant SBC in the thermodynamic limit.

Under these assumptions, a node belongs to the giant SBC if it has at least two incoming links and two outgoing links connected to the giant in-component and giant out-component, respectively.
Equivalently, we exclude nodes that either (i) are not connected to 
the giant in- and out-components, or 
(ii) are connected to these components through only a single in-link or out-link \cite{newman2008bicomponents}.
By expressing this condition through generating functions, the size $B$ of the giant SBC is given by:
\begin{align}
B &= \sum_{k_i,k_o} P(k_i,k_o) \left[ 1 - u^{k_i} - (1-u)k_i u^{k_i-1} \right] \nonumber \\
    &\quad \times \left[ 1 - v^{k_o} - (1-v)k_o v^{k_o-1} \right].
\end{align}
Here, $(1-u)k_i u^{k_i-1}$ and $(1-v)k_o v^{k_o-1}$
represent the probabilities that a node connects to the giant in- and out-components,
respectively, through exactly one link, while the remaining $k_i-1$ and $k_o-1$
links do not reach these components. 
Using the generating functions defined earlier, we can express the size $B$
of the giant SBC as:
\begin{align}
B &= 1 + G(u,v) + (1-u)(1-v)\partial_{x}\partial_{y} G(u,v) \nonumber \\
&\quad - G(u,1) - (1-u)\partial_x [G(u,1) - G(u,v)] \nonumber \\
&\quad - G(1,v) - (1-v)\partial_y [G(1,v) - G(u,v)].
\end{align}
where $\partial_x G(u,v)$ denotes $\left.\frac{\partial G(x,y)}{\partial x}\right|_{x=u,y=v}$.
The expression for the size $B$ of the giant SBC simplifies when there is no correlation between in- and out-degrees,
meaning that $P(k_i,k_o)=P(k_i)P(k_o)$:
\begin{align}
B = \left[1-G(u,1)-(1-u)\partial_x G(u,1) \right] \nonumber \\
\times \left[1-G(1,v)-(1-v)\partial_y G(1,v)\right].
\end{align}
Furthermore, for uncorrelated networks with identical in- and out-degree distributions, $u=v$ and the equation reduces to:
\begin{align}
B = \left[1-G(u,1)-(1-u)\partial_x G(u,1) \right]^2.
\end{align}

\subsection{Percolation threshold}

The percolation threshold at which the giant SBC 
first emerges can be identified by the Jacobian matrix $\mathbf{J}$ of Eqs.~\ref{eq:uv}. 
The trivial fixed point $(u,v)=(1,1)$ of Eqs.~\ref{eq:uv} loses stability when the largest eigenvalue of $\mathbf{J}$
exceeds unity, indicating the onset of a nontrivial solution that corresponds to the emergence of the giant SBC. 
Linearizing Eqs.~\ref{eq:uv} around the trivial fixed point, the condition for the emergence of the giant SBC where the point loses its stability,
\begin{align}
\frac{\langle k_i k_o\rangle}{\langle k \rangle} = 1,
\end{align}
which is identical to the condition for the giant SCC \cite{dorogovtsev2001giant}. 
This coincidence reveals a relationship between the two components: 
the giant SBC and the giant SCC emerge simultaneously at the same 
critical point. 
This result reflects the structural constraint that a biconnected component can only exist as part of a strongly connected component.

\subsection{Robustness to random failure}

Real-world networks are inevitably subject to random failures 
of nodes and links.
Understanding how the giant SBC responds to such failures is essential for measuring the structural robustness of SBC.
To this end, we analyze the percolation behavior of the giant SBC under  node or link removal.
In site percolation, each node is independently active with probability $f$
and removed with probability $1-f$ 
whereas in bond percolation, each link is independently activated with probability $f$ and removed with probability $1-f$.
In both cases, the self-consistency equations are modified to account for the possibility that a neighbor or the connection to a neighbor is absent.
Let $m$ and $n$ denote the probabilities that a randomly chosen in-link or out-link fails to connect 
to the giant in- or out-component, respectively:
\begin{align}
m &= (1-f) + f\, G_i(m,1), \\
n &= (1-f) + f\, G_o(1,n),
\end{align}
where the first term represents the probability that a node or link is removed and the second term represents the probability that the corresponding node or link is present but does not lead to the giant in- or out-component.
Therefore the self-consistency equations for the probabilities $m$ and $n$ are the same for site and bond percolation models.
In both cases, the percolation threshold $f_c$ is determined by linear stability analysis, which yields
\begin{align}
f_c=\frac{\langle k \rangle}{\langle k_i k_o\rangle},
\end{align}
which is identical for the giant SBC and SCC.

The expressions for the sizes of the giant SCC and SBC are different between site and bond percolation.
The sizes of the giant SCC and SBC for site percolation are given by 
\begin{align}
S &= f \left[ 1 - G(m,1) - G(1,n)+ G(m,n) \right],\\
B &= f \Big\{ 1 + G(m,n)+ (1-m)(1-n)\partial_{x} \partial_{y} G(m,n) \nonumber \\
&\quad - G(m,1) -(1-m)\partial_x [G(m,1) - G(m,n)] \nonumber \\
&\quad - G(1,n) -(1-n)\partial_y [G(1,n) - G(m,n)] \Big\}.
\end{align}
For bond percolation, since all nodes remain present, the factor $f$ in Eqs. (17) and (18) is absent. 
Therefore, the sizes of the giant SCC and SBC for bond percolation are given by
\begin{align}
S &=  1 - G(m,1) - G(1,n)+ G(m,n) ,\\
B &=  1 + G(m,n)+ (1-m)(1-n)\partial_{x} \partial_{y} G(m,n) \nonumber \\
&\quad - G(m,1) -(1-m)\partial_x [G(m,1) - G(m,n)] \nonumber \\
&\quad - G(1,n) -(1-n)\partial_y [G(1,n) - G(m,n)].
\end{align}
Note that the probabilities $m$ and $n$ are still determined by Eqs. (14,15) with the bond occupation probability $f$.

\section{Results}

\subsection{On random networks}

\begin{figure}
\includegraphics[width=\linewidth]{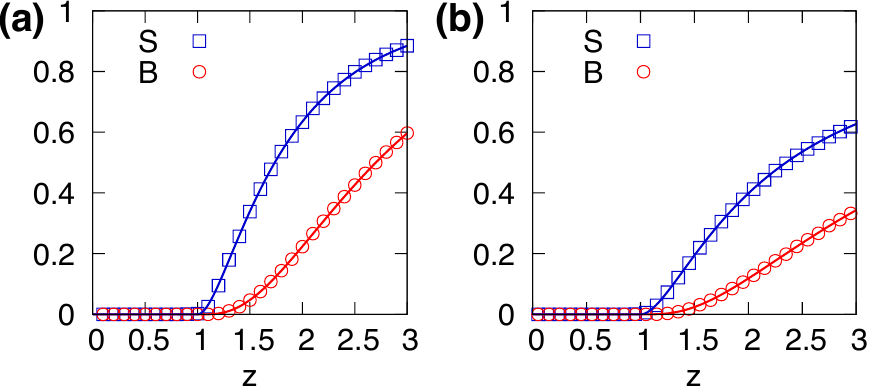}
\caption{
Size of the giant SCC and SBC in (a) directed ER networks and (b) directed networks with a Poisson in-degree and a geometric out-degree distribution, as a function of the average degree $z$ with $N=10^5$.
}
\label{fig:fig2}
\end{figure}

We first apply our theoretical framework to directed random networks 
to verify the analytical theory and to explore how the giant SBC emerges. 
We consider two representative cases: 
(i) networks where in- and out-degree distributions follow a Poisson distribution, 
and (ii) networks with a Poisson in-degree distribution and a geometric out-degree distribution. 
In both cases, the in- and out-degrees are assumed to be uncorrelated.

To identify the giant SBC numerically, we first identify the giant SCC and its strong articulation points (SAPs) \cite{georgiadis2018,georgiadis2016}, whose removal increases the number of SCCs.
We initially consider the entire giant SCC as a SBC candidate.
For each SAP, we remove the node from the original giant SCC and identify all the resulting SCCs.
We then add the removed SAP to each SCC and compute its intersections with the current SBC candidates.
These intersections are used to update the SBC candidates.
We repeat this procedure for every SAP and identify the largest remaining candidate as the giant SBC.
We use this procedure to identify the giant SBC throughout this work.

For directed Erd\H{o}s--R\'enyi (ER) random graphs, both in- and out-degrees follow a Poisson distribution with mean degree $z$. 
The corresponding generating function takes the simple form
\begin{align}
G(x,y) = e^{z(x+y-2)},
\end{align}
and the self-consistency equations for $u$ and $v$ reduce to
\begin{align}
u = e^{z(u-1)}, \qquad v = e^{z(v-1)}.
\end{align}
Since the in- and out-degree distributions are identical, $u=v$, 
and the sizes of the giant SCC and SBC are given by
\begin{align}
S &= \left( 1 - u \right)^2, \\
B &= \left[ 1 - u- z(1 - u)u \right]^2.
\end{align}
Figure~\ref{fig:fig2}(a) presents the results for both numerical simulations and analytical predictions, 
showing good agreement.
The giant SBC emerges at the same percolation threshold as the giant SCC, but its size increases more slowly 
due to the stricter connectivity requirements.

We next examine the effect of asymmetric in- and out-degree distributions. 
Specifically, we consider a network with a Poisson in-degree distribution and 
a geometric out-degree distribution, $P(k_o) = \frac{1}{z+1} \left(\frac{z}{z+1}\right)^{k_o}$, both with mean degree $z$. 
The corresponding generating function for the out-degree distribution is 
\begin{align}
G_o(1,v) = \frac{1}{z+1-z v}.
\end{align}
In this case, the generating functions for the in- and out-degree directions 
are no longer identical, and the self-consistency equations are given by 
\begin{align}
u = e^{z(u-1)}, \qquad v = \frac{1}{z}.
\end{align}
The sizes of the giant SCC and SBC then follow the expressions with these distinct solutions $u$ and $v$:
\begin{align}
S &= \left(1 - u\right)(1-v), \\
B &= \left[1 - u- z(1-u)u \right](1-v)^2.
\end{align}
Both the giant SCC and SBC emerge at the same percolation threshold. 
However, beyond this threshold, the giant SBC grows more slowly than the giant SCC, as shown in Fig.~\ref{fig:fig2}(b).

\subsection{Tolerance to random failures}

\begin{figure}
\includegraphics[width=\linewidth]{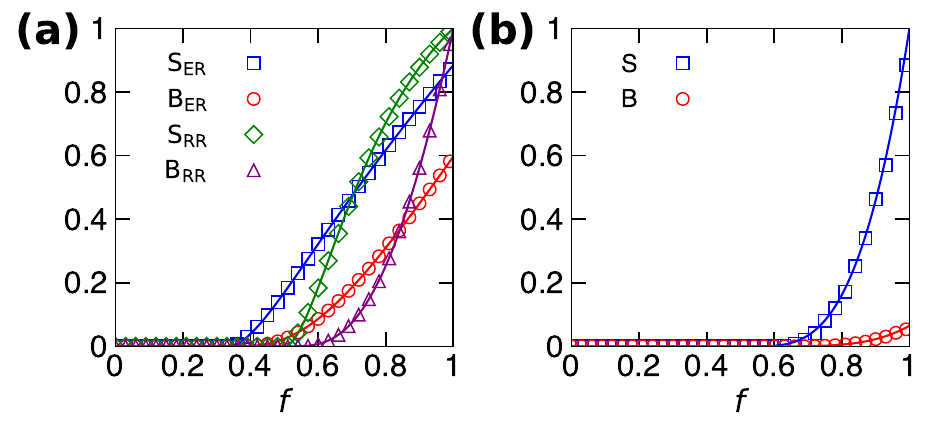}
\caption{
Size of the giant strongly biconnected and strongly connected component 
in (a) directed ER networks with degree $z=3$ and random regular networks with degree $z=2$
and (b) directed SF networks with degree exponent $\gamma=2.5$
as a function of node activation probability $f$ with $N=10^5$
averaged over $10^2$.
}
\label{fig:fig3}
\end{figure}

We next examine how the giant SBC responds to random node removal. 
Assuming site percolation, each node is independently removed with probability $1-f$ 
and activated with probability $f$.
For directed ER networks with mean in- and out-degree $z$,
the self-consistency equations become
\begin{align}
m = 1 - f + f e^{z(m-1)},
\end{align}
with $m=n$. 
The sizes of the giant SCC and SBC are then given by
\begin{align}
S &= f \left[1 - e^{z(m-1)}\right]^2, \\
B &= f \left[1 - e^{z(m-1)} - z(1-m)e^{z(m-1)}\right]^2.
\end{align}
The percolation threshold is then $f_c = 1/z$, 
at which both the giant SCC and SBC emerge simultaneously, as shown in Fig.~\ref{fig:fig3}(a).

Next, we consider directed random regular graphs where every node has fixed in- and out-degrees equal to $z$ as shown in Fig.~\ref{fig:fig3}(a).
For the specific case $z = 2$, the self-consistency equation is given by $m = 1 - f + f m^2$, 
and the percolation threshold is $f_c=1/2$.
The probability $m$ can be expressed explicitly as: 
\begin{align}
m = \begin{cases} 
1 & \text{if } f \le f_c,\\ 
\frac{1 - f}{f} & \text{if } f > f_c. 
\end{cases}
\end{align}
Substituting this into the general expressions gives the sizes of the giant SCC and SBC as:
\begin{align}
S &= f \left(1 - m^2 \right)^2, \\
B &= f \left(1 - m \right)^4.
\end{align}
A critical behavior near the percolation threshold $f \to f_c$ shows  
that the giant SCC and SBC exhibit distinct behaviors, scaling as $S \sim (f - f_c)^2$ 
and $B \sim (f - f_c)^4$, respectively \cite{luczak2009critical}.

We next consider directed scale-free (SF) networks with in- and out-degree distributions as \cite{kryven2017general}:
\begin{align}
P(k_i,k_o)=\frac{k_i^{-\gamma}k_o^{-\gamma}}{\zeta^2(\gamma)},
\end{align}
where $\gamma>2$ and $k_i,k_o\geq 1$.
The mean in- and out-degree is given by
$z=\zeta(\gamma-1)/\zeta(\gamma)$.
The corresponding generating function can be expressed in terms of the polylogarithm $\mathrm{Li}_{\gamma}(x)$ as
\begin{align}
G(x,y)=\frac{\mathrm{Li}_{\gamma}(x)\mathrm{Li}_{\gamma}(y)}{\zeta^2(\gamma)}.
\end{align}
Since the in- and out-degree distributions are independent and identical, the self-consistency equations reduce to
\begin{align}
m=1-f+f\frac{\mathrm{Li}_{\gamma}(m)}{\zeta(\gamma)}.
\end{align}
The sizes of the giant SCC and SBC are then given by
\begin{align}
S &= f\left[1-\frac{\mathrm{Li}_{\gamma}(m)}
{\zeta(\gamma)}\right]^2,\\
B &= f\left[1-\frac{\mathrm{Li}_{\gamma}(m)}
{\zeta(\gamma)}
-\frac{(1-m)\mathrm{Li}_{\gamma-1}(m)}
{m\zeta(\gamma)}\right]^2.
\end{align}
The percolation threshold is given by
$f_c=\zeta(\gamma)/\zeta(\gamma-1)$,
which is identical for the giant SCC and SBC.
The sizes of $S$ and $B$ on the SF networks with $\gamma=2.5$ are shown in Fig.~\ref{fig:fig3}(b), showing great agreement between the theoretical predictions and numerical results.

\subsection{Behaviors near the percolation threshold}

We study the critical behavior of the giant SCC and SBC near the percolation threshold.
We define the critical exponents $\beta_S$ and $\beta_B$ as
\begin{align}
S\sim(f-f_c)^{\beta_S}, \quad B\sim(f-f_c)^{\beta_B}.
\end{align}
For directed random networks with a finite second moment of the degree distribution, 
such as ER networks, we introduce $\epsilon=1-m$ and expand the generating function as
\begin{align}
G(1-\epsilon,1)\simeq 1-z\epsilon+C\epsilon^2,
\end{align}
where $C=\langle k_i(k_i-1)\rangle/2$.
Then, the self-consistency equation of $m$ gives the relation $\epsilon \sim (f-f_c)$.
We then obtain $S\sim\epsilon^2$ and $B\sim\epsilon^4$, leading to the mean-field exponents $\beta_S=2$ and $\beta_B=4$.

For directed SF networks, the critical behavior depends on the degree exponent $\gamma$.
For $2<\gamma < 3$, the second moment diverges, and the generating function shows a nonanalytic expansion,
\begin{align}
G(1-\epsilon,1)\simeq 1-z\epsilon+A\epsilon^{\gamma-1},
\end{align}
where $A$ is a constant.
The self-consistency equation then gives
$\epsilon\sim(f-f_c)^{1/(\gamma-2)}$.
Since the leading linear terms in the expression for $B$ cancel, 
the critical exponents for $2<\gamma < 3$ are given by
\begin{align}
\beta_S=\frac{2}{\gamma-2}, \quad
\beta_B=\frac{2(\gamma-1)}{\gamma-2}.
\end{align}
For $\gamma>3$, we recover the mean-field exponents
$\beta_S=2$ and $\beta_B=4$.
These results show that the critical exponent of the giant SBC is distinct from that of the giant SCC, and is not necessarily twice that of the giant SCC.

\subsection{Applications to real-world biological networks}

We finally study the application of our theory to empirical biological networks.
Here, our generating function approach is a degree-based null model for real-world networks. 
We preserve the joint in- and out-degree distribution while neglect other structural correlations.
Therefore, agreement with the theoretical predictions indicates that the degree distribution provides a good approximation of the network connectivity, whereas deviations may reflect additional structural properties not captured by the model.

\begin{figure}
\includegraphics[width=\linewidth]{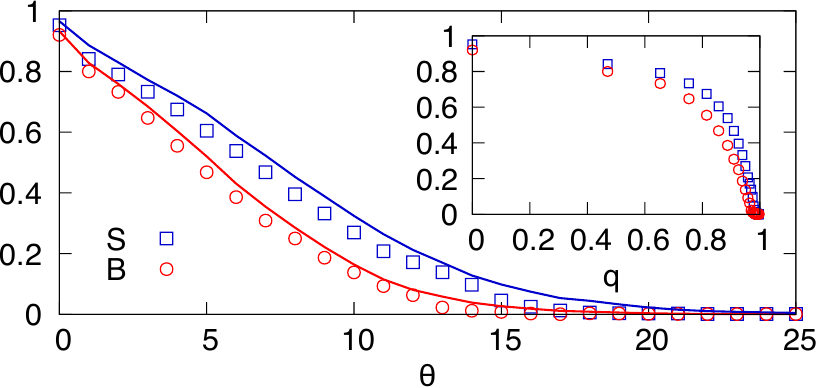}
\caption{
Size of the giant strongly biconnected and strongly connected component 
in a network of neural connectome of the larval brain as a function 
of the cutoff threshold $\theta$.
Inset shows the sizes of $S$ and $B$ as a function of the fraction $q$
of removed edges.
}
\label{fig:fig4}
\end{figure}

We apply our framework to a neural connectome of the larval brain~\cite{Michael2023connectome}. 
In this dataset, there are $N=2,952$ nodes and $L=110,677$ edges with non-zero weights, $w_{ij}$. 
This network is both directed and weighted, where each weight represents the number of synapses between two neurons.
We perform a thresholding procedure analogous to bond percolation: 
links with weight $w_{ij} \le \theta$ are removed, and the sizes of the giant SCC and SBC 
are computed as functions of the cut-off $\theta$ \cite{restrepo2008weighted}.
Figure~\ref{fig:fig4} shows the sizes of the giant SCC and SBC as a function of 
the weight cut-off $\theta$, with the theoretical predictions.
The theoretical predictions were obtained by measuring the joint in- and out-degree distribution $P(k_i,k_o)$ from the empirical network with the threshold $\theta$ and using it to construct the corresponding generating functions. 
The sizes of the giant SCC and SBC were then calculated by numerically solving the self-consistency equations. 
The theoretical curves agree the empirical results well, 
with the size of the giant SBC consistently smaller than that of the giant SCC.
Inset of Fig.~\ref{fig:fig4} shows the size of giant SCC and SBC as a function of the fraction $q$ of removed links.
These results show that our analytical framework provides reliable predictions for the connectivity structure 
of a real-world neural network under link removal.

\begin{figure}[t]
\includegraphics[width=\linewidth]{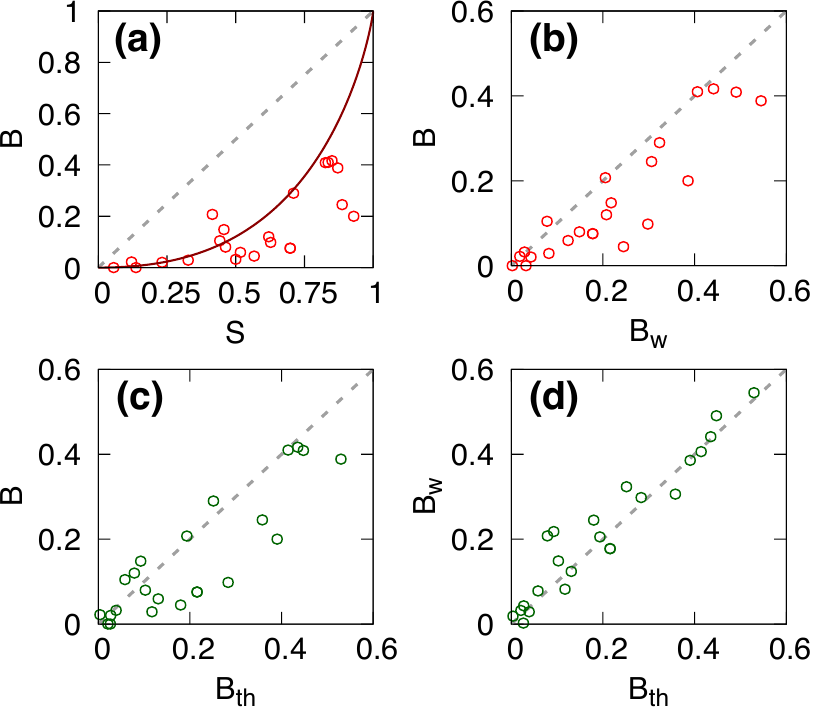}
\caption{
Giant SBC size in Cell Collective networks gathered from \cite{helikar2012cell}.
(a) $B$ vs.\ $S$ for each network; solid line is the theoretical prediction for directed ER graphs. 
(b) $B$ vs.\ $B_w$ for degree-preserving rewired networks, 
where $\langle B_w - B \rangle = 0.0656 \pm 0.0137$ (mean $\pm$ standard error).
(c) $B$ vs.\ theoretical prediction $B_{th}$ from the empirical joint degree distribution $P(k_i,k_o)$, 
and (d) $B_w$ vs.\ $B_{th}$. 
The correlation coefficients are $r=0.878$ and $r=0.957$, respectively.
}
\label{fig:fig5}
\end{figure}

We also analyze the directed networks gathered in the Cell Collective database \cite{helikar2012cell}, 
which includes gene regulatory and signaling networks across diverse biological systems. 
We use 14 directed networks whose size is larger than $50$, see the Table 1.
For each network, we compute the empirical sizes of the giant SCC and SBC. 
Figure~\ref{fig:fig5}(a) shows the comparison between $B$ and $S$ for each network, 
with the solid line representing the theoretical relationship derived for directed ER graphs,
indicating that biological networks exhibit smaller giant SBC than expected from ER networks.
The theoretical predictions were based on the joint in- and out- degree distribution $P(k_i,k_o)$ obtained from each empirical network. 
We then construct the generating functions and compute the sizes of $B$ and $S$.

We also compare the real-world networks with rewired randomized networks that preserve the exact degree sequence of the empirical networks. 
The randomized networks are generated using the degree-preserving edge-swap procedure: at each step, two directed edges, $(i \rightarrow j)$
and $(k \rightarrow l)$, are chosen at random and rewired to $(i \rightarrow l)$
and $(k \rightarrow j)$, provided that neither the resulting self-loops nor multi-edges are created. 
Because this operation only exchanges the target nodes of the two edges, the out-degrees of $i$ and $k$ and the in-degrees of $j$ and $l$ are preserved.
Then, we measure the size of the giant SBC of the rewired networks, which is denoted as $B_w$.
As shown in Fig.~\ref{fig:fig5}(b), the randomized networks exhibit a larger giant SBC than their real-world biological counterparts as $\langle B_w -B \rangle = 0.0656 \pm 0.0136$. 
These results suggest that structural properties beyond the joint in- and out-degree distribution may affect the size of the giant SBC in real biological networks
\cite{milo2002network,alon2007network,ravasz2002hierarchical,kim2015biconnectivity,ahn2006wiring}.
One possible way of the organization is that feedback loops and mutual reachability tend to be concentrated within localized functional modules rather than being distributed globally to form a large biconnected component, with redundant paths. 

\begin{table*}[t!]
\centering
\caption{Real-world Boolean networks \cite{helikar2012cell} }
\begin{tabular}{ccccccccccc}
\hline \hline
Boolean networks & $N$  & $B$ & $B_w$ & $B_{th}$ & $S$ & $z$ & $\sigma_{in}$ & $\sigma_{out}$\\
\hline
B bronchiseptica \& T retortaeformis & 53 & 0.245 & 0.306 & 0.354 & 0.887 & 2.55 & 1.60 & 2.00 \\ 
\hline
CD4 T cell signaling & 188 & 0.080 & 0.149 & 0.158 & 0.463 & 1.87 & 1.74 & 1.70 \\ 
\hline
Colitis-associated colon cancer & 70 & 0.200 & 0.386 & 0.412 & 0.929 & 2.19 & 1.07 & 1.51\\ 
\hline
ErbB (1-4) Receptor Signaling & 247 & 0.409 & 0.491 & 0.499 & 0.826 & 4.47 & 3.88 & 5.46\\ 
\hline
HIV-1 interactions w/ T Cell Signaling & 138 & 0.290 & 0.323 & 0.336 & 0.710 & 2.67 & 2.34 & 3.02 \\ 
\hline
IL-1 Signaling & 118 & 0.059 & 0.124 & 0.121 & 0.517 & 1.85 & 1.20 & 1.90 \\ 
\hline
IL-6 Signaling & 86 & 0.105 & 0.078 & 0.085 & 0.442 & 1.73 & 1.40 & 2.06 \\ 
\hline
Immune System Model & 164 & 0.207 & 0.205 & 0.229 & 0.415 & 2.61 & 2.41 & 2.51  \\ 
\hline
Lymphopoiesis Regulatory Network & 81 & 0.148 & 0.218 & 0.226 & 0.457 & 1.98 & 1.74 & 1.77 \\ 
\hline
MAPK Cancer Cell Fate Network & 53 & 0.075 & 0.178 & 0.211 & 0.698 & 1.96 & 1.26 & 1.57\\ 
\hline
PC12 Cell Differentiation & 62 & 0.032 & 0.029 & 0.019 & 0.500 & 1.74 & 0.88 & 3.37 \\ 
\hline
Signal Transduction in Fibroblasts & 139 & 0.388 & 0.545 & 0.550 & 0.871 & 4.01 & 2.64 & 3.68 \\ 
\hline
Signaling in Macrophage Activation & 321 & 0 & 0.002 & 0.002 & 0.056 & 1.66 & 1.43 & 4.94 \\ 
\hline
T-LGL Survival Network 2011 & 60 & 0.417 & 0.441 & 0.452 & 0.850 & 3.25 & 1.73 & 6.77 \\ 
\hline
\end{tabular}
\end{table*}

To further test our theoretical framework, we compare both $B$ and $B_w$, 
with respect to the theoretical prediction $B_{th}$ obtained 
by using the generating function formalism with empirical joint degree distributions $P(k_i,k_o)$.
Figure~\ref{fig:fig5}(c) shows a linear relationship between $B$ and $B_{th}$ with correlation coefficient $r=0.878$, 
indicating that our approach reproduces the size of the giant SBC with reasonable accuracy even for the small networks in the Cell Collective database. 
In addition, $B_w$ correlates even more tightly with $B_{th}$ with a higher 
coefficient $r=0.957$ as shown in Fig.~\ref{fig:fig5}(d). 
Rewiring destroys the local correlations and non-tree-like motifs present in real biological networks, leading to the randomized networks closer to the locally tree-like structure assumed by the theory. 
These results show that real-world biological networks are systematically less redundant than random graphs, implying additional local structure such as  hierarchical and modular organization.
We note that the biological networks examined in this study are relatively small, and thus finite-size effects cannot be completely ruled out. 
Further investigations using larger biological networks are needed to determine whether the observed tendency is a general feature of biological networks.

\section{Discussion}

In this work, we developed a generating function framework to compute the size of the giant SBC and its percolation behavior under random node and link removal. 
We applied it to both random and real-world networks and confirm our theory.
Our results show that directed bicomponent provide a more stringent measure of structural core and the robustness in directed networks than SCC. 
Although both emerge at the same percolation threshold, the giant SBC grows more slowly due to its requirement of multiple paths, confirming that the giant SBC captures the failure-tolerant cores that sustain connectivity under perturbations. 
Many natural systems, including the brain \cite{Michael2023connectome,Schlegel2024whole}, gene regulatory circuits \cite{helikar2012cell}, food webs \cite{garlaschelli2004patterns}, and climate networks \cite{deza2015assessing}, are intrinsically directed, and the cycles within them can be a source of stability. 
The SBC framework offers a simple way to characterize the redundancy that sustains such cycles. Future work can extend this approach to study 
how the structural patterns shape functional stability in real-world complex systems.

\begin{acknowledgments}
This research was supported in part by the National
Research Foundation of Korea (NRF) grant funded by
the Korea government (MSIT) (No. RS-2025-25433094),
by Global - Learning \& Academic research institution for Master’s · PhD students, and Postdocs (LAMP) Program of the National Research Foundation of Korea (NRF) grant funded by the Ministry of Education (No. RS-2024-00445180),
and by the IITP(Institute of Information \& Communications Technology Planning
\& Evaluation)-ITRC(Information Technology Research
Center) grant funded by the Korea government(Ministry
of Science and ICT) (IITP-2025-RS-2024-00437284).
\end{acknowledgments}

\bibliography{percolation}

\end{document}